# Emoji Retrieval from Gibberish or Garbled Social Media Text: A Novel Methodology and A Case Study

Shuqi Cui, Nirmalya Thakur, and Audrey Poon

Department of Computer Science, Emory University, Atlanta, GA 30322, USA

nicole.cui@emory.edu
nirmalyathakur@ieee.org
audrey.poon@emory.edu

**Abstract**: Emojis, considered an integral aspect of social media conversations, are widely used on almost all social media platforms. However, social media data may be noisy and may also include gibberish or garbled text which is difficult to detect and work with. Most naïve data preprocessing approaches recommend removing such gibberish or garbled text from social media posts before performing any form of data analysis or before passing such data to any machine learning model. However, it is important to note that such gibberish or garbled text may have been an emoji(s) in the original social media post(s) and failure to retrieve the actual emoji(s) may result in the loss or lack of contextual meaning of the analyzed social media data. The work presented in this paper aims to address this challenge by proposing a three-step reverse engineering-based novel methodology for retrieving emojis from garbled or gibberish text in social media posts. The development of this methodology also helped to unravel the reasons that could lead to the generation of gibberish or garbled text related to data mining of social media posts. To evaluate the effectiveness of the proposed methodology, the model was applied to a dataset of 509,248 Tweets about the Mpox outbreak, that has been used in about 30 prior works in this field, none of which were able to retrieve the emojis in the original Tweets from the gibberish text present in this dataset. Using our methodology, we were able to retrieve a total of 157,748 emojis present in 76,914 Tweets in this dataset by processing the gibberish or garbled text. The effectiveness of this methodology has been discussed in the paper through the presentation of multiple metrics related to text readability and text coherence which include the Flesch Reading Ease, Flesch Kincaid Grade Score, Coleman Liau index, Automated Readability Index, Dale Chall Readability Score, Text Standard, and Reading Time for the Tweets before and after the application of the methodology to the Tweets. The results showed that the application of this methodology to the Tweets improved the readability and coherence scores. Finally, as a case study, the frequency of emoji usage in these Tweets about the Mpox outbreak was analyzed and the results are presented.

## 1. Introduction

The emergence of social media platforms coupled with their ubiquitousness and interconnectedness has resulted in a tremendous increase in the amount of time individuals have been spending on social media in the last decade and a half [1]. On average, a person spends 143 minutes per day on social media [2] and owns 8.4 social media accounts [3]. On a global scale, about 5 billion people use social media platforms [4]. This extensive adoption of social media results in the generation of tremendous amounts of Big Data. Mining and analysis of this Big Data has attracted the attention of researchers from different disciplines for various use cases such as understanding, interpreting, anticipating, and predicting the patterns of social media discourse, sentiment analysis, subjectivity analysis, opinion mining, topic modeling, misinformation detection, and toxicity analysis, just to name a few [5-8]. Emojis, considered an integral aspect of social media conversations, are widely used on almost all social media platforms [9]. For example, on Facebook Messenger more than 900 million emojis are sent every day, half of all the comments on Instagram include an emoji, and one in five Tweets includes an emoji [10, 11]. However, social media data is noisy [12] and prior works in this field have also stated that the issues with social media data include gibberish or garbled text [13,14] which is difficult to detect [15]. In this context, we define "gibberish" or "garbled" text as text that contains characters outside the character set of all human-readable languages and such characters cannot be comprehended by humans. Table 1 represents a few Tweets that contain garbled or gibberish text for further clarification of the definition.

| **Table 1**: Examples of Tweets that contain gibberish or garbled text |
|---|
| **Tweet #1**: Actually the K is silent. It is pronounced MONEYPOX ðŸ• µðŸ™ŠðŸ™ˆðŸ'°ðŸ'°ðŸ'°ðŸ'°ðŸ'°ðŸ'°ðŸ'°ðŸ'°ðŸ'°ðŸ'µðŸ'´ðŸ'°ðŸ'´ðŸ'µðŸ'´ µðŸ'´ðŸ'¶ðŸ'¶ðŸ'°ðŸ'·ðŸ'µðŸ'´ðŸ'´ðŸ'´ðŸ'´ðŸ'µðŸ'°ðŸ'°ðŸ'°ðŸ'°ðŸ'°ðŸ'°ðŸ'°ðŸ'°ðŸ'°ðŸ'°ðŸ'°ðŸ'°ðŸ'°ðŸ'°ðŸ'°ðŸ'µðŸ'µðŸ'µðŸ'µðŸ'µðŸ'°ðŸ'°ðŸ'µðŸ'µðŸ'° |
| **Tweet #2**: @MaureenStroud @MatthewNewell67 @nuhope2022 @handmadekathy @JimeeLiberty @volpiranyas @pawley_robert @PremChamp1 @PLHartungRN @FvckYourFear @Cdoglover1 @BlueBear0386 @doom37455413 @provaxtexan @The_Aussie_Luke @GhostDancer2022 @TsuDhoNimh @ogilville1 @AngryFleas @BadC19TestTakes @LizNYC13 @mcfunny @Chelle389 @SkepticalMutant @swedishchf @KStateTurk @itisjustmebabe @YellowstoneRan1 @Nockit1 @andylumm @KathyGa28615606 @richykirsh @theanswer50 @HighJanky @GeoffSchuler @NoMisinfoToday @ConsequentialBr @JonDaley7 @TonyBaduy @noonienoodie @doritmi @sammy44231 @zeetubes @JonathanHannah @Monstercoyliar @tomsirolimus @SallyJiggles @AndrewLazarus4 @butterednoodIe It seems that every few weeks they have something new and ridiculous. The whole monkey pox thing that someone tried to pin on the vaccines was just..ðŸ˜³ðŸ™„ðŸ¤¡ðŸ˜µâ€¢ ðŸ'«ðŸ˜µâ€¢ ðŸ'«ðŸ˜µâ€¢ ðŸ'«ðŸ˜µâ€¢ ðŸ'«ðŸ˜µâ€¢ ðŸ'« |
| **Tweet #3**: Now they got something called monkeypox that's out ðŸ¤¦ðŸ• ¾â€¢ â™€ï'• ðŸ¤¦ðŸ• ¾â€¢ â™€ï'• ðŸ¤¦ðŸ• ¾â€¢ â™€ï'• ðŸ¤¦ðŸ• ¾â€¢ â™€ï'• ðŸ¤¦ðŸ• ¾â€¢ â™€ï'• ðŸ¤¦ðŸ• ¾â€¢ â™€ï'• |
| **Tweet #4:** Monkey pox whispers in work. Anyone fooled by this fear a second time ðŸ¤¦â€¢ â™€ï'• ðŸ¤¦â€¢ â™€ï'• ðŸ¤¦â€¢ â™€ï'• ðŸ¤¦â€¢ â™€ï'• ðŸ¤¦â€¢ â™€ï'• ðŸ¤¦â€¢ â™€ï'• ðŸ¤¦â€¢ â™€ï'• ðŸ¤¦â€¢ â™€ï'• |
| **Tweet #5**: @GBNEWS Of course he would ðŸ™„is his relevance waning like the fear of the rona has cue monkey |

> poxðŸ'¤ðŸ'¤ðŸ'¤ðŸ'¤ðŸ'¤ðŸ'¤ðŸ'¤ðŸ'¤ðŸ'¤ðŸ'¤ðŸ¤®ðŸ¤®ðŸ¤®ðŸ¤®ðŸ¤®ðŸ¤®ðŸ¤®ðŸ¤®ðŸ¤®ðŸ'‰ðŸ'‰ðŸ'‰ðŸ'‰ðŸ'‰ðŸ'‰ðŸ'‰ðŸ'‰ðŸ'‰ðŸ'‰ðŸ'‰ðŸ'‰

These Tweets are presented here in "as is" form after obtaining the same from the Tweet IDs of the used dataset. These Tweets do not represent or reflect the views, opinions, beliefs, or political stances of the authors of this paper.

In the last few years, there have been multiple works related to the detection or prediction of emojis in social media texts (reviewed in Section 2). The ongoing and past Mpox outbreaks in different parts of the world [16-19] and conversations about the same on social media platforms have also been investigated in multiple research projects in this field (reviewed in Section 2). However, the methodologies proposed in prior works in this field for the prediction or detection of emojis are not highly accurate as none of those algorithms identify the root cause that led to the generation of a gibberish character(s) instead of an emoji in the mined version of a social media post. Furthermore, none of the prior works related to the analysis of social media posts about the ongoing and past Mpox outbreaks have focused on the retrieval of emojis from gibberish or garbled text. Addressing this research challenge serves as the main motivation for this work. The rest of this paper is organized as follows. In Section 2, a review of recent works in this field is presented. Section 3 discusses the methodology which is followed by the results in Section 4. Section 4 is followed by the conclusion section which concludes the paper and summarizes the scope of future work in this field.

## 2. Literature Review

The work of Chouhan et al. [20] showed that data gathered from live-streaming platforms frequently involves emojis, emotes, and emoticons, and understanding the usage of the same is vital in comprehending the discussions on live-streaming platforms. Kumar et al. [21] utilized machine learning and deep learning algorithms to predict emojis in text-based data where the emojis were not mined correctly or were missing. Nusrat et al. [22] used Bert to predict the most appropriate emoji for a given text. Their model achieved an overall accuracy of 75% and was more accurate as compared to other emoji prediction models at that time. The work of Kone et al. [23] concluded that emojis are a "visual language" that enables users to communicate their feelings. Ranjan et al. [24] studied the relationship between English words and emojis to predict the latter. They trained their model using concepts of Multinomial Naive Bayes and LSTMs to predict emojis in Tweets. Stoikos et al. [25] utilized a BERTmoticon model to predict missing emojis in social media texts. They applied the model to Tweets about COVID-19. The findings of their work showed that since WHO's declaration of COVID-19 as a global pandemic, there has been a spike in the usage of emojis on social media platforms, specifically emojis that are associated with feelings of sadness. Inan et al. [26] used concepts of classification from machine learning to predict emojis in Tweets. They tested their model on a dataset of Tweets and their approach achieved a 0.901 F1 score. Kumar et al. [27] proposed an on-device pipeline to insert emojis in appropriate locations in texts based on concepts of semantic analy-sis. Barbieri et al. [28] developed a methodology that used concepts of word embedding in vectorial space to predict emojis in social media texts. A similar approach was used by Gupta et al. [29]. However, the work of Gupta et al. [29] also incorporated the time and location information associated with social media texts to improve the accuracy of emoji predictions. Their methodology achieved an overall accuracy of 73.32%. Shobana et al. [30] developed a deep neural network for predicting emojis in texts where the underlying emojis were missing or not mined correctly. Their methodology used concepts of text semantic analysis and achieved an overall accuracy of 86%. Barbieri et al. [31] studied the semantics behind emojis and developed a model for the prediction of emojis in tweets. Zhao et al. [32] developed a dataset of Emoji-embedded Tweets and post-response pairs to study their learning method for persona-aware emoji-embedded dialogue.

Sv et al. [33] studied public attitudes towards the Mpox outbreak. They performed sentiment analysis on 556,403 English Tweets about the Mpox outbreak, published between June 1, 2022, and June 25, 2022. The findings showed that the percentage of neutral, positive, and negative tweets was 41.6%, 28.82%, and 23.01%, respectively. A similar study of sentiment analysis of Tweets about the Mpox outbreak was performed by Ng et al. [34]. They used 352,182 tweets published between May 6, 2022, and July 23, 2022. Cooper et al. [35] analyzed Tweets about Mpox published between May 1, 2022, and July 23, 2022. The findings of their work showed that 48,330 tweets were written by LGBTQ+ individuals or advocates and the most common sentiment present in these Tweets was fear or sadness. Iparraguirre-Villanueva et al. [36] developed a sentiment analysis model using CNN and LSTM to perform sentiment analysis of Tweets about Mpox. Their model achieved an overall accuracy of 83%. D'souza et al. [37] studied 70,832 Tweets published between May 01, 2022, and September 07, 2022, that contained the terms #monkeypox, #MPVS, #stigma, or #LGBTQ+. The results of their study showed that the LGBTQ+ community faced hate on Twitter as a result of stigma, misinformation, and misinterpretation related to the Mpox outbreak. Zuhanda et al. [38] analyzed a dataset of 5000 Tweets about Mpox published on August 5, 2022. The findings showed that fear was the most prevalent emotion expressed in the Tweets. Knudsen et al. [39] studied Tweets about Mpox published between May 18, 2022, and September 19, 2022. The findings of their work showed that 82% of the Tweets expressed incorrect information about Mpox. Bengesi et al. [40] studied Tweets about Mpox for sentiment analysis. They proposed a methodology that used TextBlob annotation, Lemmatization, Vectorization, and a Support Vector Machine for performing sentiment analysis. Their approach achieved an overall accuracy of 93.48%. Farahat et al. [41] performed sentiment analysis of 8532 Tweets about Mpox published between May 22, 2022, and August 5, 2022. The findings of their work showed that the percentage of neutral, positive, and negative tweets was 48%, 37%, and 15%, respectively.

In summary, even though multiple research works have been conducted in this field thus far, two major research gaps still remain:

- Emojis may be represented as garbled or gibberish text during the data mining process of posts (for example: Tweets) from social media platforms. The process of predicting emojis to replace gibberish text has achieved considerable at-tention from the scientific community in the last few years. However, the developed algorithms are not highly accurate as they don't investigate or identify the root cause that led to the generation of a gibberish character(s) instead of an emoji(s) in the mined version of a social media post. So, the predicted emoji(s) by the existing algorithms may not always be correct. As emojis are a crucial aspect of communication on social media platforms, incorrect emojis may completely change the meaning of a sentence. For example, if we consider these two statements – "Did your friend really die? 😂" and "Did your friend really die? 😭". The first sentence contains the "face with tears of joy" emoji. It is an emoji that represents crying with laughter facial expression. The second statement contains the "loudly crying face" emoji. This emoji depicts a face with an open mouth and streams of heavy tears flowing from closed eyes. In this context, between the two emojis, the "loudly crying face" emoji would be appropriate. However, if an algorithm (incorrectly) predicts the "face with tears of joy" emoji at the end of this sentence, it would be inappropriate for the given context.

- Sentiment analysis of Tweets continues to attract the attention of researchers from different disciplines. Considering emojis is crucial for performing sentiment analysis [42-47]. Most naïve data preprocessing approaches recommend removing gibberish or garbled texts from social media posts (for example: Tweets) before performing any form of data analysis or before passing such data to any machine learning model [48,49]. While it is true that the gibberish or garbled text would have a negative effect on the data analysis task or on the training of any machine learning algorithm (not just for sentiment analysis), it is important to note that the gibberish text or garbled text contains meaning (for instance, it may represent one or more emojis) which would get lost if such data was directly deleted. Prior works in this field, for example [50], have followed such naïve data preprocessing approaches i.e. ignoring the garbled or gibberish text in social media content prior to data analysis and machine learning-based model development.

Addressing these gaps with an aim to contribute to the advancement of research in this field, serves as the main motivation for this research work. A detailed discussion of the methodology is presented in Section 3.

## 3. Methodology

This section presents the methodology that was developed to address the research challenge of retrieving emojis from gibberish or garbled text by proposing a three-step reverse engineering-based novel methodology. The first step of this approach uses concepts of Natural Language Processing and Information Retrieval to detect garbled or gibberish text in social media posts. The second step uses concepts of Reverse Engineering, Data Analysis, and Information Processing to infer the underlying cause for the generation of the garbled or gibberish text in a social media post. The final step reverse engineers the process to restore the emojis as were present in the original post on a social media platform. A detailed discussion of the working of this methodology is presented later in this section. The flowchart shown in Figure 1 presents an overview of this methodology as well as an overview of the research work and the case study that was performed.

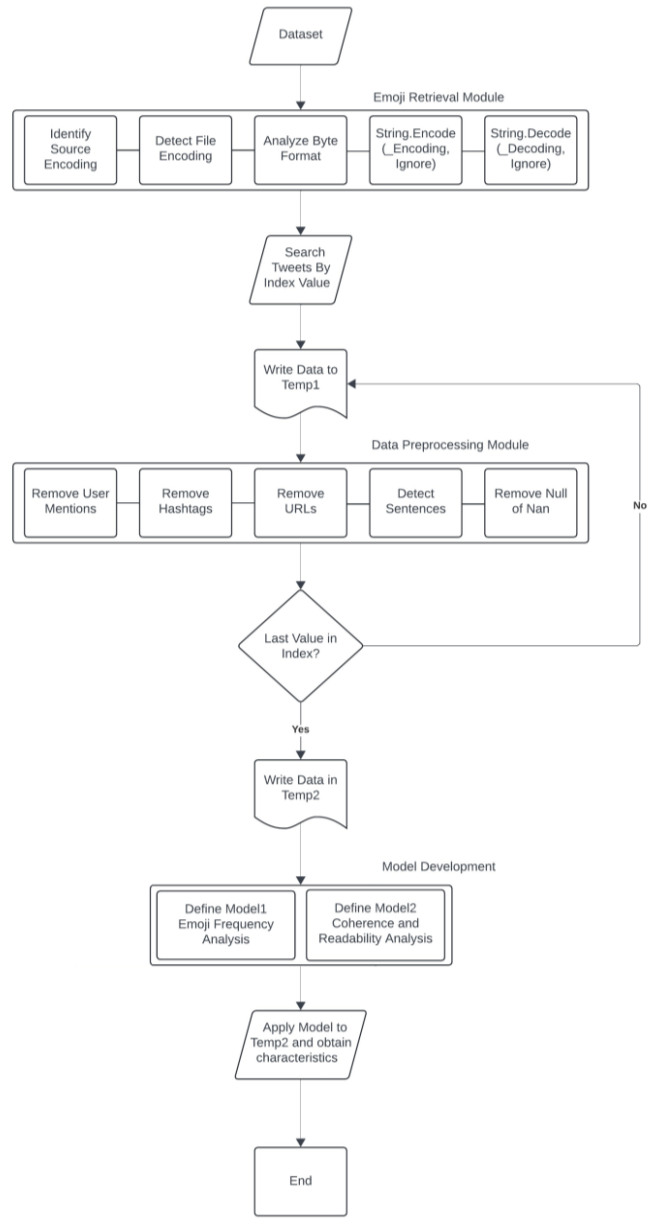

**Figure 1**: A flowchart that represents an overview of the methodology and the specifics of the case study that was performed.

The development of this methodology also helped to unravel the reasons that could lead to the generation of gibberish or garbled text related to data mining of social media text. The work showed that gibberish or garbled text may be generated by computer programs or algorithms during the data mining process from a social media platform due to one or more of the following scenarios:

- Text from social media platforms is decoded in a format different from the source text during the data mining process.
- Mismatch of default encoding across applications or platforms or programs during the data analysis process, or machine learning algorithm development process where social media data was used.
- The application of incorrect encoding for storage or retrieval of data files containing social media data.

To evaluate the effectiveness of the proposed reverse engineering-based methodology for retrieving emojis from garbled or gibberish text, we wrote a Python program to implement and apply the same to a dataset of 509,248 Tweets about the Mpox outbreak published on Twitter between May 21, 2022, to November 11, 2022 [51]. This dataset was specifically selected for the evaluation of this methodology for multiple reasons. First, several Tweets in this dataset contain gibberish or garbled text. Second, even though this dataset has been used in about 30 prior works involving data analysis of Tweets related to the Mpox outbreak, for example [52-55], none of those works focused on retrieving emojis from the gibberish or garbled text which is present in several Tweets in this dataset. Third, different parts of the world are currently experiencing an outbreak of Mpox including the United States where the number of cases in 2024 is nearly double as compared to the number of cases of Mpox in the United States around the same time last year [56,57].

As shown in Figure 1, the Emoji Retrieval Module contains multiple functions that are associated with specific tasks. These tasks include identifying source encoding, detecting file encoding, analyzing byte format, performing string encoding, and performing string decoding. In this context, the source encoding refers to the encoding used in the character set (for example: English) used to publish content on a social media platform (for example: Twitter) that belongs to the list of encoding patterns accepted by that platform. Here, "file encoding" refers to the type of encoding in the output file(s) by a data mining program, algorithm, or software to retrieve that content from that social media platform. Upon identification of any garbled or gibberish text, the second step of our methodology uses a brute force approach to identify the exact cause of the garbled or gibberish text and the exact encoding patterns that represent the values of "source encoding" and "file encoding". More specifically, in this step, the algorithm iterates through a list of all encoding patterns in Python 3.12 to identify the value of the "file encoding". Table 2 shows these codecs by name and the languages for which the encoding is likely used [58]. For the development of this approach, the encodings supported by Python 3.12 were used for two reasons. First, Python 3.12 was the most recent version of Python at the time of development of this methodology. Second, the list of encodings supported by Python 3.12 is more than the list of encodings supported by old versions of Python [59-63]. Upon identification of the "file encoding", our algorithm uses a reverse engineering-based methodology to identify the source encoding. More specifically, it processes emojis using every encoding-to-encoding conversion using the value of the "file encoding" to determine the value of the "source encoding". Thereafter, using the values of the "source encoding" and "file encoding" it reverse engineers the process to retrieve the gibberish or garbled text to its original form that represented one or more emojis. The remainder of Figure 1 shows the steps that we performed for the case study on this dataset to evaluate the effectiveness of our methodology. The results that highlight the effectiveness of this methodology are presented in Section 4.

| Table 2: Encodings supported by Python 3.12 | |
|---|---|
| Codec | Languages |
| ascii | English |
| big5 | Traditional Chinese |
| big5hkscs | Traditional Chinese |
| cp037 | English |
| cp273 | German |
| cp424 | Hebrew |
| cp437 | English |
| cp500 | Western Europe |
| cp720 | Arabic |
| cp737 | Greek |
| cp775 | Baltic languages |
| cp850 | Western Europe |
| cp852 | Central and Eastern Europe |
| cp855 | Bulgarian, Byelorussian, Macedonian, Russian, Serbian |
| cp856 | Hebrew |
| cp857 | Turkish |
| cp858 | Western Europe |
| cp860 | Portuguese |
| cp861 | Icelandic |
| cp862 | Hebrew |
| cp863 | Canadian |
| cp864 | Arabic |
| cp865 | Danish, Norwegian |
| cp866 | Russian |
| cp869 | Greek |
| cp874 | Thai |
| cp875 | Greek |
| cp932 | Japanese |
| cp949 | Korean |
| cp950 | Traditional Chinese |
| cp1006 | Urdu |
| cp1026 | Turkish |
| cp1125 | Ukrainian |
| cp1140 | Western Europe |
| cp1250 | Central and Eastern Europe |
| cp1251 | Bulgarian, Byelorussian, Macedonian, Russian, Serbian |
| cp1252 | Western Europe |
| cp1253 | Greek |
| cp1254 | Turkish |
| cp1255 | Hebrew |
| cp1256 | Arabic |
| cp1257 | Baltic languages |
| cp1258 | Vietnamese |
| euc_jp | Japanese |
| euc_jis_2004 | Japanese |
| euc_jisx0213 | Japanese |
| euc_kr | Korean |
| gb2312 | Simplified Chinese |
| gbk | Unified Chinese |
| gb18030 | Unified Chinese |
| hz | Simplified Chinese |
| iso2022_jp | Japanese |
| iso2022_jp_1 | Japanese |
| iso2022_jp_2 | Japanese, Korean, Simplified Chinese, Western Europe, Greek |
| iso2022_jp_2004 | Japanese |
| iso2022_jp_3 | Japanese |
| iso2022_jp_ext | Japanese |
| iso2022_kr | Korean |
| latin_1 | Western Europe |
| iso8859_2 | Central and Eastern Europe |
| iso8859_3 | Esperanto, Maltese |
| iso8859_4 | Baltic languages |
| iso8859_5 | Bulgarian, Byelorussian, Macedonian, Russian, Serbian |
| iso8859_6 | Arabic |
| iso8859_7 | Greek |
| iso8859_8 | Hebrew |
| iso8859_9 | Turkish |
| iso8859_10 | Nordic languages |
| iso8859_11 | Thai languages |
| iso8859_13 | Baltic languages |
| iso8859_14 | Celtic languages |
| iso8859_15 | Western Europe |
| iso8859_16 | South-Eastern Europe |
| johab | Korean |
| koi8_r | Russian |
| koi8_t | Tajik |
| koi8_u | Ukrainian |
| kz1048 | Kazakh |
| mac_cyrillic | Bulgarian, Byelorussian, Macedonian, Russian, Serbian |
| mac_greek | Greek |
| mac_iceland | Icelandic |
| mac_latin2 | Central and Eastern Europe |
| mac_roman | Western Europe |
| mac_turkish | Turkish |
| ptcp154 | Kazakh |
| shift_jis | Japanese |
| shift_jis_2004 | Japanese |
| shift_jisx0213 | Japanese |
| utf_32 | all languages |
| utf_32_be | all languages |
| utf_32_le | all languages |
| utf_16 | all languages |
| utf_16_be | all languages |
| utf_16_le | all languages |
| utf_7 | all languages |
| utf_8 | all languages |
| utf_8_sig | all languages |

## 4. Results and Discussions

This section presents the results of this research work. To show the effectiveness of the proposed three-step reverse engineering-based novel methodology for retrieving emojis from garbled or gibberish text in social media posts, we have presented Table 3. In Table 3, we revisit Table 1 and show the processing of those Tweets using our methodology that results in the retrieval of all the emojis (from the gibberish or garbled text) as were present in these original Tweets on Twitter.

| |
|---|
| **Table 3**: Results of applying our methodology to the Tweets from Table 1 |
| **Tweet #1 from Table 1 (before applying our methodology)**: Actually the K is silent. It is pronounced MONEYPOX<br>ðŸ• µðŸ™ŠðŸ™ˆðŸ'°ðŸ'°ðŸ'°ðŸ'°ðŸ'°ðŸ'°ðŸ'°ðŸ'°ðŸ'µðŸ'´ðŸ'°ðŸ'°ðŸ'´ðŸ'µðŸ'µðŸ'´ðŸ'¶ðŸ'¶ðŸ'°ðŸ'·ðŸ'µðŸ'´ðŸ'´ðŸ'´ðŸ'´ðŸ'µðŸ'°ðŸ'°ðŸ'°ðŸ'°ðŸ'°ðŸ'°ðŸ'°ðŸ'°ðŸ'°ðŸ'°ðŸ'°ðŸ'°ðŸ'µðŸ'µðŸ'µðŸ'µðŸ'µðŸ'°ðŸ'°ðŸ'µðŸ'µðŸ'° |
| **Tweet #1, from Table 1, after processing the same using our methodology**:<br>Actually the K is silent It is pronounced MONEYPOX<br>🙊🙈💰💰💰💰💰💰💰💰💵💴💰💰💴💵💵💴💶💶💰💷💵💴💴💴💴💵💰💰💰💰💰💰💰💰💰💰💰💰💵💵💵💵💵💰💰💵💵💰 |
| **Tweet #2 from Table 1 (before applying our methodology)**: @MaureenStroud @MatthewNewell67 @nuhope2022 @handmadekathy @JimeeLiberty @volpiranyas @pawley_robert @PremChamp1 @PLHartungRN @FvckYourFear @Cdoglover1 @BlueBear0386 @doom37455413 @provaxtexan @The_Aussie_Luke @GhostDancer2022 @TsuDhoNimh @ogilville1 @AngryFleas @BadC19TestTakes @LizNYC13 @mcfunny @Chelle389 @SkepticalMutant @swedishchf @KStateTurk @itisjustmebabe @YellowstoneRan1 @Nockit1 @andylumm @KathyGa28615606 @richykirsh @theanswer50 @HighJanky @GeoffSchuler @NoMisinfoToday @ConsequentialBr @JonDaley7 @TonyBaduy @noonienoodie @doritmi @sammy44231 @zeetubes @JonathanHannah @Monstercoyliar @tomsirolimus @SallyJiggles @AndrewLazarus4 @butterednoodIe It seems that every few weeks they have something new and ridiculous. The whole monkey pox thing that someone tried to pin on the vaccines was just..ðŸ˜³ðŸ™„ðŸ¤¡ðŸ˜â€¢ ðŸ'«ðŸ˜µâ€¢ ðŸ'«ðŸ˜µâ€¢ ðŸ'«ðŸ˜µâ€¢ ðŸ'«ðŸ˜µâ€¢ ðŸ'« |
| **Tweet #2, from Table 1, after processing the same using our methodology:**<br>@MaureenStroud @MatthewNewell67 @nuhope2022 @handmadekathy @JimeeLiberty @volpiranyas @pawley_robert @PremChamp1 @PLHartungRN @FvckYourFear @Cdoglover1 @BlueBear0386 @doom37455413 @provaxtexan @The_Aussie_Luke @GhostDancer2022 @TsuDhoNimh @ogilville1 @AngryFleas @BadC19TestTakes @LizNYC13 @mcfunny @Chelle389 @SkepticalMutant @swedishchf @KStateTurk @itisjustmebabe @YellowstoneRan1 @Nockit1 @andylumm @KathyGa28615606 @richykirsh @theanswer50 @HighJanky @GeoffSchuler @NoMisinfoToday @ConsequentialBr @JonDaley7 @TonyBaduy @noonienoodie @doritmi @sammy44231 @zeetubes @JonathanHannah @Monstercoyliar @tomsirolimus @SallyJiggles @AndrewLazarus4 @butterednoodIe It seems that every few weeks they have something new and ridiculous. The whole monkey pox thing that someone tried to pin on the vaccines was just😳🙄🤡😵💫😵💫😵💫😵💫 |
| **Tweet #3 from Table 1 (before applying our methodology)**: Now they got something called monkeypox that's out<br>ðŸ¤¦ðŸ• ¾â€• â™€ï¸• ðŸ¤¦ðŸ• ¾â€• â™€ï¸• ðŸ¤¦ðŸ• ¾â€• â™€ï¸• ðŸ¤¦ðŸ• ¾â€• â™€ï¸• ðŸ¤¦ðŸ• ¾â€• â™€ï¸• ðŸ¤¦ðŸ• ¾â€• â™€ï¸• |
| **Tweet #3, from Table 1, after processing the same using our methodology:**<br>Now they got something called monkeypox thats out 🙍‍♀🙍‍♀🙍‍♀🙍‍♀🙍‍♀ |
| **Tweet #4 from Table 1 (before applying our methodology)**: Monkey pox whispers in work. Anyone fooled by this fear a second time<br>ðŸ¤¦â€• â™€ï¸• ðŸ¤¦â€• â™€ï¸• ðŸ¤¦â€• â™€ï¸• ðŸ¤¦â€• â™€ï¸• ðŸ¤¦â€• â™€ï¸• ðŸ¤¦â€• â™€ï¸• ðŸ¤¦â€• â™€ï¸• ðŸ¤¦â€• â™€ï¸• |
| **Tweet #4, from Table 1, after processing the same using our methodology:**<br>Monkey pox whispers in work Anyone fooled by this fear a second time<br>🙍‍♀🙍‍♀🙍‍♀🙍‍♀🙍‍♀🙍‍♀🙍‍♀ |
| **Tweet #5 from Table 1 (before applying our methodology)**: @GBNEWS Of course he would ðŸ™„is his relevance waning like the fear of the rona has cue monkey poxðŸ’¤ðŸ’¤ðŸ’¤ðŸ’¤ðŸ’¤ðŸ’¤ðŸ’¤ðŸ’¤ðŸ’¤ðŸ’¤ðŸ¤®ðŸ¤®ðŸ¤®ðŸ¤®ðŸ¤®ðŸ¤®ðŸ¤®ðŸ¤®ðŸ¤®ðŸ’‰ðŸ’‰ðŸ’‰ðŸ’‰ðŸ’‰ðŸ’‰ðŸ’‰ðŸ’‰ðŸ’‰ðŸ’‰ |
| **Tweet #5, from Table 1, after processing the same using our methodology:**<br>GBNEWS Of course he would 🙄 is his relevance waning like the fear of the rona has cue monkey<br>pox 💤💤💤💤💤💤💤💤💤💤🤮🤮🤮🤮🤮🤮🤮🤮🤮💉💉💉💉💉💉💉💉💉💉 |

These Tweets are presented here in "as is" form after obtaining the same from the Tweet IDs of the used dataset and after processing the same using the proposed methodology. These Tweets do not represent or reflect the views, opinions, beliefs, or political stances of the authors of this paper.

In a similar way, using this methodology we were able to retrieve all the emojis from the garbled or gibberish text in this dataset. There were a total of 76914 Tweets containing garbled or gibberish text and upon passing the same through our methodology we were able to retrieve the actual emojis that were used in these Tweets. A total of 157748 emojis were retrieved during this process. We evaluated the effectiveness of the application of this methodology to this dataset using multiple metrics related to text readability and text coherence which include the Flesch Reading Ease, Flesch Kincaid Grade Score, Coleman Liau index, Automated Readability Index, Dale Chall Readability Score, Text Standard, and Reading Time for the Tweets before and after the application of the methodology. Twitter allows a user to publish a Tweet containing up to 280 characters and users may or may not use all the characters. In fact, studies have shown that depending on the subject matter users on Twitter may post short or long Tweets [64,65]. Therefore, it is crucial to investigate the performance of our methodology on Tweets of different lengths (in terms of the number of characters). So, we grouped the Tweets in this dataset into four groups as outlined below:

- Group 1 = length of Tweets less than or equal to 70 characters
- Group 2 = length of Tweets greater than or equal to 71 characters but less than 141 characters

- Group 3 = length of Tweets greater than or equal to 141 characters but less than 211 characters
- Group 4 = length of Tweets greater than 211 characters

Figure 2 shows the evaluation of these metrics for all these groups before applying our methodology and Figure 3 shows the evaluation of these metrics for all these groups after applying our methodology.

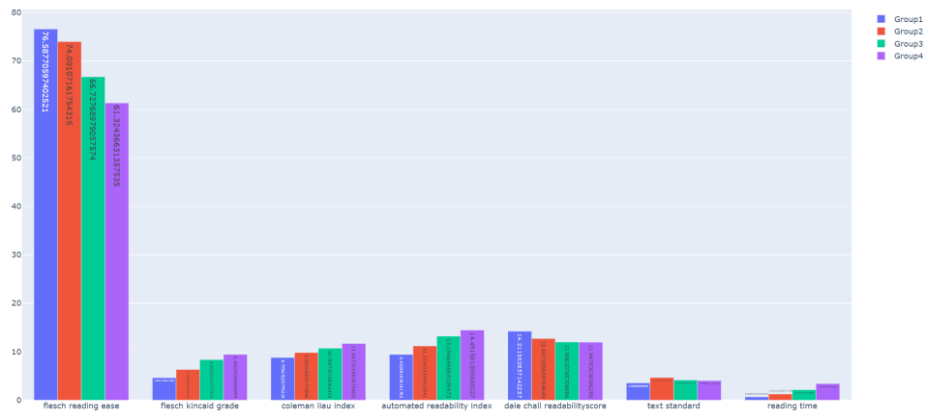

**Figure 2**: Evaluation of different metrics before applying our methodology

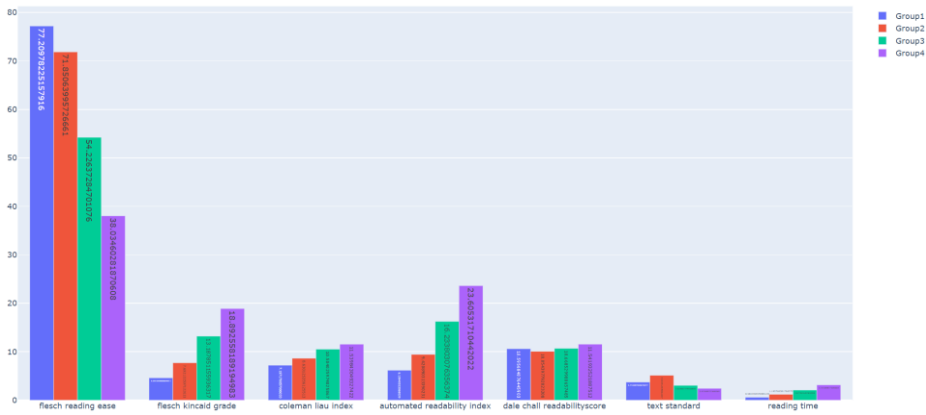

**Figure 3**: Evaluation of different metrics after applying our methodology

The improvement of most of these metrics upon the application of our methodology indicates the effectiveness of the same related to text readability and coherence. For instance, for Group 1, the average Automated Readability Index improved from 9.41 to 6.15. The average Automated Readability Index is a number that approximates the grade level needed to comprehend a given text. In other words, for this metric, a lower score indicates better readability. Similarly, for Group 1, the average Dale Chall Readability Score changed from 14.21 to 10.59.  The reading time for Group 1, Group 2, Group 3, and Group 4 before applying our methodology was 0.66, 1.24, 2.12, 3.38. After applying our methodology, the reading time for all these groups improved to 0.58, 1.19, 2.07, and 3.13, respectively.

To further support the scientific contributions of this work, we performed a frequency analysis of all the emojis present in this dataset as none of the prior works that used this dataset were able to retrieve any emoji(s) from the gibberish or garbled text present in the Tweets. The results of the same (top 40 emojis are shown for the paucity of space) are presented in Table 4.

**Table 4:** Results from performing frequency analysis of emojis (top 40 emojis are shown for the paucity of space) in the dataset

| Emoji Name | Emoji Symbol | Frequency |
|---|---|---|
| joy |  | 11328 |
| sob |  | 9120 |
| rofl |  | 6795 |
| thinking |  | 4217 |
| facepalm |  | 3038 |
| male_sign | ♂ | 2761 |
| roll_eyes |  | 2685 |
| female_sign | ♀ | 2680 |
| weary |  | 2335 |
| shrug |  | 2240 |
| woozy_face |  | 2033 |
| see_no_evil |  | 1723 |
| flushed |  | 1685 |
| clown_face |  | 1520 |
| nauseated_face |  | 1394 |
| skull |  | 1388 |
| speak_no_evil |  | 1375 |
| vomiting_face |  | 1292 |
| eyes |  | 1252 |
| syringe |  | 1212 |
| mask |  | 1079 |
| unamused |  | 1053 |
| point_down |  | 990 |
| rotating_light |  | 956 |
| hear_no_evil |  | 955 |
| grimacing |  | 933 |
| satisfied |  | 929 |
| sweat_smile |  | 923 |
| dizzy_face |  | 816 |
| dizzy |  | 776 |
| wink |  | 755 |
| 100 |  | 753 |
| upside_down_face |  | 738 |
| melting_face |  | 708 |
| microbe |  | 692 |
| tired_face |  | 627 |
| raised_eyebrow |  | 583 |
| scream |  | 577 |
| smiling_face_with_tear |  | 572 |
| zany_face |  | 556 |

The work presented in this paper has a couple of limitations. First, the second step of our methodology uses a brute force approach to identify the exact cause of the garbled or gibberish text and the exact encoding patterns that represent the values of "source encoding" and "file encoding". So, from the time complexity standpoint, there is scope for improvement in this regard which was not addressed in this paper. Second, the list of encodings supported by Python 3.12 was used for the development of this methodology as the encodings supported by Python 3.12 are more than old versions of Python. Python 3.12 was the most recent version of Python at the time of development of this methodology. However, future versions of Python may support a higher number of encodings. As any new encoding patterns that future versions of Python may support were not stated on python.org [66] at the time of development of this methodology, this probable limitation could not be addressed at this point.

## 5. Conclusion

Emojis, an essential component of social media posts, are extensively used on almost all social media platforms. However, social media data is prone to noise, and prior works in this area have also acknowledged the presence of gibberish text in social media posts, that is difficult to identify and work with. In the recent past, multiple studies have focused on identifying and predicting the presence of emojis in social media posts containing gibberish text. However, those algorithms are not highly accurate as those algorithms do not investigate or identify the root cause for the generation of gibberish text in the mined version of the social media posts. Therefore, the emojis generated by these algorithms may not consistently align with the intended meaning of an analyzed social media post(s). Furthermore, emojis play a vital role in the development of various Natural Language Processing algorithms. Conventional data preprocessing approaches often advise the deletion of such gibberish or garbled text from social media posts, such as Tweets, prior to performing any data analysis or providing such posts as input to a machine learning model. In this context, it is crucial to note that such gibberish or garbled text may contain meaningful information, such as emojis, which would be lost if simply deleted.

The work presented in this paper aims to address this challenge by proposing a three-step reverse engineering-based novel methodology for retrieving emojis from garbled or gibberish text in social media posts. The development of this methodology also helped to unravel the reasons that could lead to the generation of gibberish or garbled text related to data mining of social media text. To evaluate the effectiveness of the proposed methodology, it was applied to a dataset of 509,248 Tweets about the Mpox outbreak which has been used in several prior works related to sentiment analysis and other analyses of Tweets about Mpox. However, none of those works retrieved the emojis from the garbled or gibberish text in this dataset. Upon applying our methodology to this dataset, we were able to successfully retrieve all the emojis from the garbled or gibberish text in this dataset. There were a total of 76914 Tweets containing garbled or gibberish text and upon passing the same through our methodology we were able to retrieve the actual emojis that were used in these Tweets. A total of 157748 emojis were retrieved during this process. The effectiveness of the application of this methodology to this dataset has been discussed in the paper through the presentation of multiple metrics related to text readability and text coherence which include the Flesch Reading Ease, Flesch Kincaid Grade Score, Coleman Liau index, Automated Readability Index, Dale Chall Readability Score, Text Standard, and Reading Time for the Tweets before and after the application of the methodology. The results showed that the application of this methodology to the Tweets improved the readability and coherence scores.

As per the best knowledge of the authors, no similar work has been performed in this field thus far. Future work would involve improving the time complexity and integrating the methodology as a Python package that can be imported into Python programs for emoji retrieval from gibberish or garbled texts in social media posts.

**Disclosure of Interests.** The authors have no competing interests to declare that are relevant to the content of this article.